\documentclass[lettersize,journal]{IEEEtran}
\usepackage{amsmath,amsfonts}
\usepackage{algorithmic}
\usepackage{algorithm}
\usepackage{array}
\usepackage[caption=false,font=normalsize,labelfont=sf,textfont=sf]{subfig}
\usepackage{textcomp}
\usepackage{stfloats}
\usepackage{url}
\usepackage{verbatim}
\usepackage{graphicx}
\usepackage{cite}
\usepackage{multirow}
\usepackage{booktabs}
\usepackage{xcolor}
\usepackage{colortbl}
\usepackage{hyperref}

\usepackage{subcaption}

\newcommand{\revise}[1]{\textcolor{black}{#1}}  

\begin{document}

\title{
Wavelet-Domain Masked Image Modeling for Color-Consistent HDR Video Reconstruction
}

\author{
    Yang Zhang,
    Zhangkai Ni,~\IEEEmembership{Member,~IEEE}, 
    Wenhan Yang,~\IEEEmembership{Member,~IEEE},
    Hanli Wang,~\IEEEmembership{Senior Member,~IEEE}

\thanks{This work was supported in part by the National Natural Science Foundation of China under Grants 62571372 and 62371343, and in part by the Fundamental Research Funds for the Central Universities.}
\thanks{Yang Zhang and Zhangkai Ni are with the School of Computer Science and Technology and the Key Laboratory of Embedded System and Service Computing, Ministry of Education, Tongji University, Shanghai 200092, China (e-mail: zhangy\_ce@tongji.edu.cn; zkni@tongji.edu.cn).}
\thanks{Hanli Wang is with the College of Electronic and Information Engineering, the School of Computer Science and Technology, and the Key Laboratory of Embedded System and Service Computing, Ministry of Education, Tongji University, Shanghai 200092, China. (e-mail: hanliwang@tongji.edu.cn)}
\thanks{Wenhan Yang is with Pengcheng Laboratory, Shenzhen, Guangdong 518066, China. (e-mail: yangwh@pcl.ac.cn).}
\thanks{Corresponding author: Zhangkai Ni (email: zkni@tongji.edu.cn).}
}

\markboth{Journal of \LaTeX\ Class Files,~Vol.~14, No.~8, August~2021}%
{Shell \MakeLowercase{\textit{et al.}}: A Sample Article Using IEEEtran.cls for IEEE Journals}


\maketitle

\begin{abstract}
High Dynamic Range (HDR) video reconstruction aims to recover fine brightness, color, and details from Low Dynamic Range (LDR) videos. However, existing methods often suffer from color inaccuracies and temporal inconsistencies.
To address these challenges, we propose WMNet, a novel HDR video reconstruction network that leverages Wavelet domain Masked Image Modeling (W-MIM). 
WMNet adopts a two-phase training strategy:
In Phase I, W-MIM performs self-reconstruction pre-training by selectively masking color and detail information in the wavelet domain, enabling the network to develop robust color restoration capabilities. 
A curriculum learning scheme further refines the reconstruction process.
Phase II fine-tunes the model using the pre-trained weights to improve the final reconstruction quality.
To improve temporal consistency, we introduce the Temporal Mixture of Experts (T-MoE) module and the Dynamic Memory Module (DMM). T-MoE adaptively fuses adjacent frames to reduce flickering artifacts, while DMM captures long-range dependencies, ensuring smooth motion and preservation of fine details.
Additionally, since existing HDR video datasets lack scene-based segmentation, we reorganize HDRTV4K into HDRTV4K-Scene, establishing a new benchmark for HDR video reconstruction.
Extensive experiments demonstrate that WMNet achieves state-of-the-art performance across multiple evaluation metrics, significantly improving color fidelity, temporal coherence, and perceptual quality.
The code is available at: \url{https://github.com/eezkni/WMNet}
\end{abstract}

\begin{IEEEkeywords}
High dynamic range video reconstruction, masked image modeling, memory mechanism.
\end{IEEEkeywords}

\section{Introduction}
\IEEEPARstart{C}{ompared} with low dynamic range (LDR) videos, high dynamic range (HDR) videos offer a wider brightness range, richer colors, and more realistic details, making them ideal for applications such as advanced photography, medical imaging, and autonomous driving~\cite{kim2020jsi, xu2022fmnet}. However, most consumer-grade cameras are inherently constrained by sensor limitations and can only capture LDR videos.
Consequently, converting LDR videos into HDR videos has become a crucial research focus in computer vision.
Recent advances in HDR video reconstruction have primarily followed two technical paradigms: alternating exposure and frame-by-frame reconstruction. The alternating exposure approach~\cite{chen2021hdr,chung2023lan,xu2024hdrflow} dynamically adjusts exposure settings during video capture, enabling the simultaneous acquisition of highlight and shadow details to enhance reconstruction quality. 
In contrast, the frame-by-frame reconstruction approach~\cite{kim2019deep,chen2021new,tan2021deep, guo2023learning} treats a video as a sequence of independent frames, applying single-image HDR reconstruction techniques to process each frame separately.

Since the alternating exposure method imposes additional requirements on input data and is difficult to generalize for reconstructing arbitrary LDR videos, the frame-by-frame approach has become the predominant solution for HDR video reconstruction. Despite significant progress, it still faces three key challenges. First, it does not effectively utilize temporal information across frames, limiting reconstruction quality. Second, independent frame processing often leads to temporal inconsistencies, resulting in flickering or unstable outputs. Third, as LDR and HDR videos exist in different color spaces, accurate color space transformation is crucial.
However, existing HDR video reconstruction methods struggle to ensure color fidelity, frequently resulting in artifacts and inaccuracies in reconstructed HDR content.

\begin{figure}
\begin{center}
\includegraphics[width=1.0\linewidth]{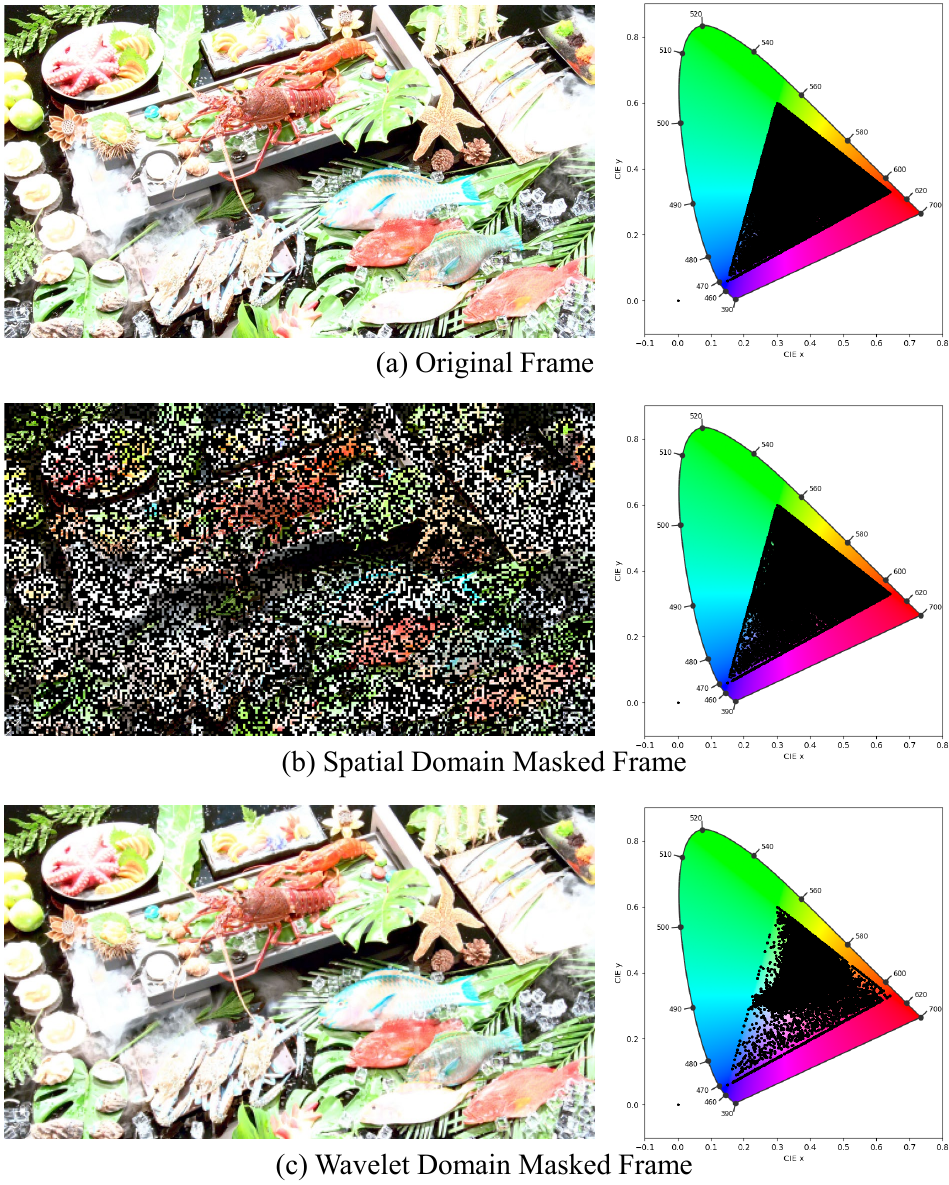}
\end{center}
\caption{
\revise{Comparison of different masking strategies on color gamut distribution.
(a) Original HDR frame and its corresponding color gamut;
(b) Spatial-domain masked HDR frame and its color gamut;
(c) Wavelet-domain masked HDR frame and its color gamut.
Zoom in for clearer visualization of color differences.}}
\label{fig:motivation}
\end{figure}

To address these challenges, we propose a Wavelet domain Masked Image Modeling (W-MIM) based network for HDR video reconstruction.
First, to mitigate color deviations in reconstructed videos, we introduce a mask modeling mechanism inspired by recent advances in masked image modeling. However, directly applying random masking to video frames—even at masking rates as high as 90\%—fails to adequately cover the color spectrum (see Fig.\ref{fig:motivation}~(b)). 
In contrast, we find that performing masked modeling in the wavelet domain can effectively cover the entire color space (Fig. \ref{fig:motivation}~(c)).
We employ W-MIM during self-reconstruction pre-training, using a curriculum learning-based masking ratio strategy to progressively enhance the model's capacity for accurate color restoration.
Second, to address temporal inconsistency, we draw on insights from recent video reconstruction research~\cite{chan2022basicvsr++, jiang2023video, zhou2024upscale, zhu2024temporal, zhu2021neuspike} that leverage inter-frame information exchange. Based on these insights, we design a Temporal Mixture of Experts (T-MoE) module, where multiple experts dynamically integrate adjacent frame information to guide the reconstruction of the current frame. 
This approach significantly improves both reconstruction quality and temporal coherence.
Third, inspired by the success of memory mechanisms in tasks such as object segmentation~\cite{oh2019video, hong2022dual}, video restoration~\cite{kim2019deep, ji2022multi, zhu2022event}, semantic segmentation~\cite{kim2022pin}, and object detection~\cite{han2022global}, we introduce a Dynamic Memory Module (DMM). DMM captures and maintains global scene information, selectively retrieving the most relevant contextual cues based on the current input to further enhance temporal consistency and improve overall reconstruction fidelity.

Finally, the commonly used HDR video reconstruction datasets, HDRTV1K and HDRTV4K, are not organized by scene, making them unsuitable for direct use in training our model. To address this issue, we reselect and reorganize the HDRTV4K dataset by segmenting video frames into distinct scenes, creating a new dataset called HDRTV4K-Scene. In this dataset, we randomly select 80\% of the scenes for training and use the remaining 20\% for testing. Additionally, since individual scenes in HDRTV4K-Scene contain relatively few frames, we collect an additional set of 10 scenes with long video sequences, called HDRTV4K-LongScene, to better evaluate our model's performance on extended video content.
In summary, this paper makes the following contributions:
\begin{itemize}
    \item 
    We propose a Wavelet domain masked image modeling (W-MIM) strategy with a curriculum-learning-based masking ratio adjustment scheme to progressively enhance the model's ability to reconstruct accurate video colors.
    
    \item 
    We design a Temporal Mixture of Experts (T-MoE) module that adaptively fuses adjacent frames, guiding the reconstruction process to improve both visual quality and temporal consistency. 
    
    \item 
    We introduce a Dynamic Memory Module (DMM) that stores scene-level information, further enhancing temporal consistency and improving overall reconstruction fidelity. Extensive experiments validate that WMNet achieves state-of-the-art performance, particularly in color fidelity and temporal stability.
\end{itemize}

The rest of this paper is organized as follows. Section~\ref{sec:related} reviews related work. Section~\ref{sec:method} describes the proposed WMNet in detail. Section~\ref{sec:results} presents the experimental results, and Section~\ref{sec:conclusion} concludes the paper.

\section{Related Works}
\label{sec:related}

\subsection {HDR Video Reconstruction}
HDR video reconstruction methods can be broadly categorized into alternating exposure and frame-by-frame reconstruction approaches. This paper focuses on the latter.
With the advancement of deep learning, Chen et al.~\cite{chen2021new} proposed a three-step HDR video reconstruction framework comprising adaptive global color mapping, local enhancement, and highlight area generation. 
Xu et al.~\cite{xu2022fmnet} introduced FMNet, a frequency-aware modulation network that enhances LDR-to-HDR conversion by adaptively adjusting contrast in the frequency domain, effectively mitigating low-frequency artifacts.
Guo et al.~\cite{guo2023learning} further addressed the issue of dim and unsaturated HDR outputs, attributing them to limitations in the training data. 
They constructed a new HDR video dataset and designed an HDR-to-LDR degradation model to improve reconstruction quality.
While deep learning-based HDR video reconstruction has made notable progress, most existing methods treat it as a frame-by-frame HDR image reconstruction problem, overlooking temporal interactions between adjacent frames.
Recently, several research works have focused on the temporal consistency problem in HDR video reconstruction. For example, Ye et al.~\cite{ye2024deep} propose a temporal cue-based method, VITM-TC, which enhances temporal consistency by propagating texture information and temporal cues from reconstructed key frames to adjacent frames. However, while VITM-TC enhances temporal coherence, it does not fully resolve challenges in accurate color recovery.

\begin{figure*}[t]
\begin{center}
\includegraphics[width=1.0\linewidth]{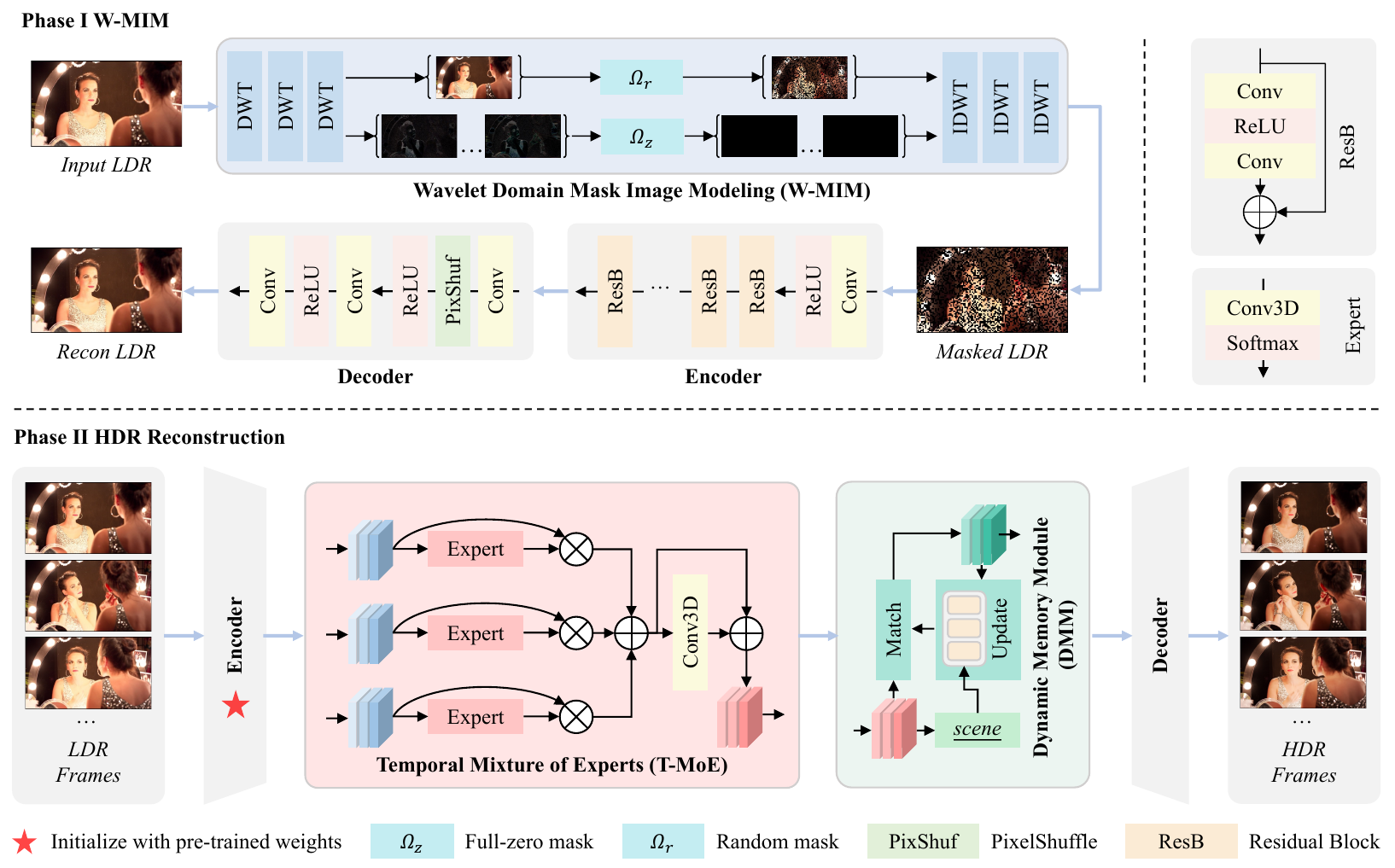}
\end{center}
\caption{\revise{Overall framework of WMNet.
Phase I performs self-reconstruction pre-training using Wavelet-domain Masked Image Modeling (W-MIM) to enhance color and detail restoration.
Phase II fine-tunes the encoder for HDR video reconstruction and incorporates the Temporal Mixture of Experts (T-MoE) and Dynamic Memory Module (DMM) to improve temporal consistency.}}
\label{fig:framework}
\end{figure*}

\subsection{Masked Image Modeling}
Self-supervised learning based on masked image modeling (MIM) has achieved remarkable progress in computer vision by enabling models to learn richer feature representations through masked region reconstruction.
Bao et al.~\cite{bao2021beit} introduced BEiT, the first MIM-based method for visual token prediction, significantly improving Vision Transformer (ViT) performance across multiple visual tasks. 
Building on this, Zhou et al.~\cite{zhou2021ibot} proposed iBOT, which integrates semantic-rich visual markers with MIM, enhancing the model’s understanding of local patterns and robustness. 
Wei et al.~\cite{wei2022masked} further advanced MIM with MaskFeat, reconstructing directional gradient histograms to improve the capture of complex spatiotemporal structures in dense visual signals.
Meanwhile, SimMIM~\cite{xie2022simmim} and MAE~\cite{he2022masked} directly reconstruct pixel values of masked patches to learn robust, generalizable representations. 
More recently, Wang et al.~\cite{wang2023masked} incorporated multi-level supervision mechanisms into MIM, strengthening the model’s ability to capture multi-scale semantic features.
Inspired by these advancements, this work explores MIM in the wavelet domain to enhance color representation learning, thereby improving HDR video reconstruction quality.

\subsection{Memory Mechanism}
Memory mechanisms have gained increasing popularity in computer vision tasks, including object segmentation, video restoration, semantic segmentation, and object tracking. Liu et al.~\cite{liu2022learning} propose a QDMN for semi-supervised video object segmentation, which selectively memorizes accurately segmented intermediate frames as references, thereby enhancing video temporal consistency. Ji et al.~\cite{ji2022multi} introduce a multi-scale memory-based deep video deblurring network that stores blurry-sharp feature pairs in a memory bank to provide useful information for blurry query inputs. Kim et al.~\cite{kim2022pin} present a memory-guided domain-generalization semantic segmentation method based on a meta-learning framework, which utilizes an external fixed memory as category guidance to reduce representation ambiguity in unseen domains. Gao et al.~\cite{gao2023memotr} propose MeMOTR, a long-term memory-enhanced Transformer for multi-object tracking, which stabilizes trajectory embeddings by incorporating long-term memory through a customized memory attention layer. These methods integrate memory mechanisms to store global information, significantly improving performance. However, existing approaches either operate within a single batch~\cite{liu2022learning,ji2022multi} or establish global memory~\cite{kim2022pin,gao2023memotr} across the entire dataset, making them vulnerable to interference from different scenes. To address this limitation, this paper introduces a scene-specific memory mechanism to enhance HDR video reconstruction more effectively.

\section{Methods}
\label{sec:method}

\subsection{Overview}
As illustrated in Figure~\ref{fig:framework}, the proposed WMNet follows a two-phase training process. \textbf{Phase I} performs self-reconstruction pre-training, while \textbf{Phase II} focuses on HDR video reconstruction.  
In Phase I, given an input video sequence \(\mathbf{x}=\{x_t\in \mathbb{R}^{H \times W \times 3}\}_{t=1}^{T}\), the proposed W-MIM strategy is applied as:  
\begin{equation}
\mathbf{x}^{\prime} = \text{M}_{\text{W-MIM}}(\mathbf{x}), 
\end{equation}
where \(\text{M}_{\text{W-MIM}}(\cdot)\) denotes the W-MIM operation, producing a masked sequence \(\mathbf{x}^{\prime}=\{x^{\prime}_{t} \in \mathbb{R}^{H \times W \times 3}\}_{t=1}^{T}\).  
Next, the masked sequence is fed into an encoder-decoder model for pre-training, aiming to reconstruct the original video:  
\begin{equation} 
\mathbf{\hat{x}} = \text{M}_{\text{DEC}}(\text{M}_{\text{ENC}}(\mathbf{x}^{\prime})),
\end{equation} 
where \(\text{M}_{\text{ENC}}(\cdot)\) and \(\text{M}_{\text{DEC}}(\cdot)\) represent the encoder and decoder modules, respectively. The reconstructed output is \(\mathbf{\hat{x}}=\{\hat{x}_{t} \in \mathbb{R}^{H \times W \times 3}\}_{t=1}^{T}\).

In Phase II, training is conducted based on the weights learned in Phase I. Given an input video sequence \(\mathbf{x}\), the encoder extracts features:  
\begin{equation}
\mathbf{z} = \text{M}_\text{ENC}(\mathbf{x}),
\end{equation}
where \(\mathbf{z} =\{F_t \in \mathbb{R}^{\frac{H}{2} \times \frac{W}{2} \times C}\}^{T}_{t=1}\) denotes the extracted features. 
The encoder \(\text{M}_{\text{ENC}}(\cdot)\) has the same structure as the encoder of Phase I, consisting of a convolutional layer for shallow feature extraction and multiple residual blocks for deep feature learning.
It is initialized with weights from Phase I to guide the model in refining details and color reconstruction.

\revise{To enhance reconstruction quality and global consistency, intermediate outputs from different residual block groups (ResB modules in Figure~\ref{fig:framework}) in the encoder are processed by T-MoE, which facilitates information exchange across video frames under the same scale through expert modules: } 
\begin{equation}
\mathbf{z^{\prime}} = \text{M}_\text{T-MoE}(\mathbf{z}^{1},\mathbf{z}^{2},...,\mathbf{z}^{d},...,\mathbf{z}^{D}),
\end{equation}
where \(D\) denotes the number of residual block groups, \(\mathbf{z}^{d}\) represents the output of the last residual block in the \(d\)-th group, and \(\mathbf{z^{\prime}}=\{F^{\prime}_t \in \mathbb{R}^{\frac{H}{2} \times \frac{W}{2} \times C}\}^{T}_{t=1}\) represents the enhanced feature after T-MoE processing.

To further improve global temporal consistency, \(\mathbf{z^{\prime}}\) is refined by the Dynamic Memory Module (DMM):  
\begin{equation}
\mathbf{\hat{z}} = \text{M}_{\text{DMM}}(\mathbf{z}^{\prime}),
\end{equation}
where \(\text{M}_{\text{DMM}}(\cdot)\) represents the DMM, and \(\mathbf{\hat{z}}=\{\hat{F}_t \in \mathbb{R}^{\frac{H}{2} \times \frac{W}{2} \times C}\}^{T}_{t=1}\) denotes the temporally enhanced feature.

Finally, the decoder reconstructs the HDR video sequence:  
\begin{equation}
\mathbf{\hat{y}} = \text{M}_{\text{DEC}}(\mathbf{\hat{z}}),
\end{equation}
where \(\text{M}_{\text{DEC}}(\cdot)\) represents the video reconstruction module, and \(\mathbf{\hat{y}}=\{\hat{y}_t \in \mathbb{R}^{H \times W \times 3}\}^{T}_{t=1}\) denotes the final HDR video output.

The following subsections provide a detailed explanation of each component and its collaborative role in achieving high-performance HDR video reconstruction.

\subsection{Wavelet Domain Mask Image Modeling (W-MIM)}
\label{subsection:W-MIM}
As illustrated in Figure~\ref{fig:motivation}, we visualize the color range under different masking approaches. 
Figure~\ref{fig:motivation} (a) shows the original video frame and its color range, which aligns with the BT.709 standard for LDR videos. Figure~\ref{fig:motivation} (b) shows the effect of random spatial-domain masking. Despite a high mask rate of 0.9 that removes substantial spatial details, the color range remains largely unchanged, suggesting that color information is not directly dependent on spatial pixels.
In contrast, Figure~\ref{fig:motivation} (c) presents the results of masking high-frequency components in the wavelet domain, where the color range is significantly reduced. 
This observation indicates that high-frequency components in the Wavelet domain play a critical role in color representation. 
Inspired by these findings, we propose a novel image modeling mechanism W-MIM that applies masking in the wavelet domain to effectively capture color and detail information in HDR video reconstruction.

Specifically, W-MIM employs a 2D discrete wavelet transform (DWT) based on the Haar wavelet~\cite{yoo2019photorealistic} to decompose the video frame into high-frequency and low-frequency components. 
DWT utilizes four convolutional kernels, denoted as \(\{\alpha \alpha^{\top}, \alpha\beta^{\top}, \beta\alpha^{\top}, \beta\beta^{\top}\}\), where \(\alpha\) represents low-pass filtering, \(\beta\) represents high-pass filtering, and \(\top\) indicates the transpose operation:
\begin{equation}
\alpha^{\top} = \frac{1}{\sqrt{2}}\left [ 1,1 \right ], \beta^{\top} = \frac{1}{\sqrt{2}}\left [ -1,1 \right ].
\end{equation}

The input video frame \( x_t \) is processed using the DWT kernels, resulting in the following decomposition:  
\begin{equation}
\revise{
\begin{split}
    \{L_{t}^{(1)}, H_{t,\alpha \beta^{\top}}^{(1)}, H_{t,\beta \alpha^{\top}}^{(1)}, H_{t,\beta \beta^{\top}}^{(1)}\}=&\text{DWT}(x_t; \alpha, \beta), \\
    \{L_{t}^{(n)}, H_{t,\alpha \beta^{\top}}^{(n)}, H_{t,\beta \alpha^{\top}}^{(n)}, H_{t,\beta \beta^{\top}}^{(n)}\}=&\text{DWT}(L_{t}^{(n-1)}; \alpha, \beta), \\
\end{split}}
\end{equation}
\revise{where \( L_{t}^{(n)} \in \mathbb{R}^{\frac{H}{2^n} \times \frac{W}{2^n} \times 3 } \) represents the low-frequency component at level \(n\), while \( H_{t,\alpha \beta^{\top}}^{(n)}, H_{t,\beta \alpha^{\top}}^{(n)}, H_{t,\beta \beta^{\top}}^{(n)} \in \mathbb{R}^{\frac{H}{2^n} \times \frac{W}{2^n} \times 3 } \) represent the corresponding high-frequency components. 
The input LDR video frames are given by \( x_t \in \mathbb{R}^{H \times W \times 3} \), where \( t=1,2,…,T \). 
In our work, we perform three levels of DWT to obtain multi-scale frequency representations.}

A full-zero mask \( \varOmega_z \) is applied to the high-frequency components, effectively discarding high-frequency information and encouraging the model to focus on restoring color and details during self-reconstruction. Meanwhile, a random mask \( \varOmega_r \) is applied to the low-frequency component to improve scene understanding and enhance reconstruction quality. The masked components are then recombined via IDWT to reconstruct the video frame \( x^{\prime}_t \) as:
\begin{equation}
\revise{
\begin{split}
    \hat{L}_t^{(N)} =& \varOmega_r \circ {L}_t^{(N)} ,\\
    \hat{H}_t^{(n)} =& \varOmega_z \circ {H}_t^{(n)} ,\\
    \hat{L}_t^{(n-1)} =& \text{IDWT}(\hat{L}_{t}^{(n)}, \hat{H}_{t,\alpha \beta^{\top}}^{(n)}, \hat{H}_{t,\beta \alpha^{\top}}^{(n)}, \hat{H}_{t,\beta \beta^{\top}}^{(n)}; \alpha, \beta), \\ 
    x_t' =& \text{IDWT}(\hat{L}_{t}^{(1)}, \hat{H}_{t,\alpha \beta^{\top}}^{(1)}, \hat{H}_{t,\beta \alpha^{\top}}^{(1)}, \hat{H}_{t,\beta \beta^{\top}}^{(1)}; \alpha, \beta),  \\
\end{split}}
\end{equation}
\revise{where \( \circ \) denotes the Hadamard product, \( N \) is the total number of DWT levels, and \( \varOmega_r \) represents a \( 1 \times 1 \) masking operator. 
To further enhance the model’s reconstruction capability, we adopt a curriculum learning strategy that gradually increases the masking ratio of low-frequency components from 0 to 0.5 during training, allowing the network to progressively learn more challenging color and detail restoration.
}

The masked video frames are then fed into a self-reconstruction pre-training model. 
As shown in Figure~\ref{fig:framework}, the model follows an encoder-decoder architecture with an asymmetric design, where the encoder is deeper than the decoder to capture richer information. 
The encoder begins with a convolutional layer for shallow feature extraction, followed by stacked residual blocks (ResB) for feature enhancement.
Each residual block consists of two convolutional layers with a ReLU activation in between. 
The decoder reconstructs the video frame by first restoring the spatial size using a convolutional layer and PixelShuffle with ReLU activation, then applying two more convolutional layers with a ReLU in between to recover the original number of channels. 
After self-reconstruction training, only the encoder’s weights are retained for the second training phase, while the decoder’s weights are discarded.

\subsection{Temporal Mixture of Experts (T-MoE)}
Videos inherently contain redundant and complementary information across adjacent frames. 
Prior studies in video super-resolution~\cite{liang2024vrt,zhou2024upscale} have demonstrated that effectively utilizing this information enhances both frame reconstruction quality and temporal consistency.  
Building on this insight, we propose the T-MoE module to facilitate inter-frame interaction. 
As illustrated in Figure~\ref{fig:framework}, we employ 3D convolution to capture temporal dependencies. 
Given that the encoder consists of stacked residual blocks, where each block extracts features at different scales, fusing these multi-scale features can further improve reconstruction. 
To achieve this, we incorporate a Mixture-of-Experts (MoE) mechanism to adaptively integrate information from various depths.

Specifically, residual blocks of the encoder are partitioned into \( D \) groups, with the output from the last residual block in each group serving as the input to the T-MoE. T-MoE is an adaptation of the existing Mixture-of-Experts (MoE) approach~\cite{pavlitska2023sparsely}, where each input is assigned to a dedicated expert. Each expert comprises a 3D convolutional layer followed by a softmax activation function, which determines the weight of the input feature by incorporating information from adjacent frames. The weighted summation of these features is then performed to obtain enhanced representations. The entire process can be formulated as follows:
\begin{equation} 
\begin{aligned}
    \bar{\mathbf{z}} &=  {\textstyle \sum_{d=1}^{D}} \mathbf{z}^{d}\circ\mathbf{z}^d_w, \\
    \mathbf{z}^d_w &=  \text{Softmax}(\text{Conv3D}(\mathbf{z}^{d})).
\end{aligned}
\end{equation}

Finally, a residual 3D convolution is applied to further refine the features, producing the final enhanced representation:
\begin{equation} 
    \hat{\mathbf{z}} = \text{Conv3D}(\bar{\mathbf{z}}) + \bar{\mathbf{z}}.
\end{equation}

In our experiments, we set \( D = 3 \) and use a 3D convolution kernel with a temporal dimension of 3 to achieve a balance between performance and efficiency.

\subsection{Dynamic Memory Module (DMM)}
In video processing, computational constraints often limit each batch to a fixed number of frames, restricting the receptive field to only a few consecutive frames. However, critical information may reside in frames outside these limited frames. To address this, existing methods~\cite{kim2022pin, gao2023memotr, xiong2024rethinking} incorporate memory mechanisms to improve global information utilization, yielding promising results. 
However, these methods typically suffer from two key limitations: (1) memory is often confined to a single batch, leading to a restricted receptive field, or (2) memory is established across the entire dataset, making it susceptible to interference from different scenes. To overcome these issues, we propose a scene-specific memory mechanism, DMM, which dynamically stores and retrieves global information within each scene, thereby enhancing the reconstruction of the entire video more effectively. 
As illustrated in Figure~\ref{fig:memory}, DMM consists of two main components: \textit{memory matching} and \textit{memory updating}.

\subsubsection{Memory Matching}
\begin{figure}
\begin{center}
\includegraphics[width=1.0\linewidth]{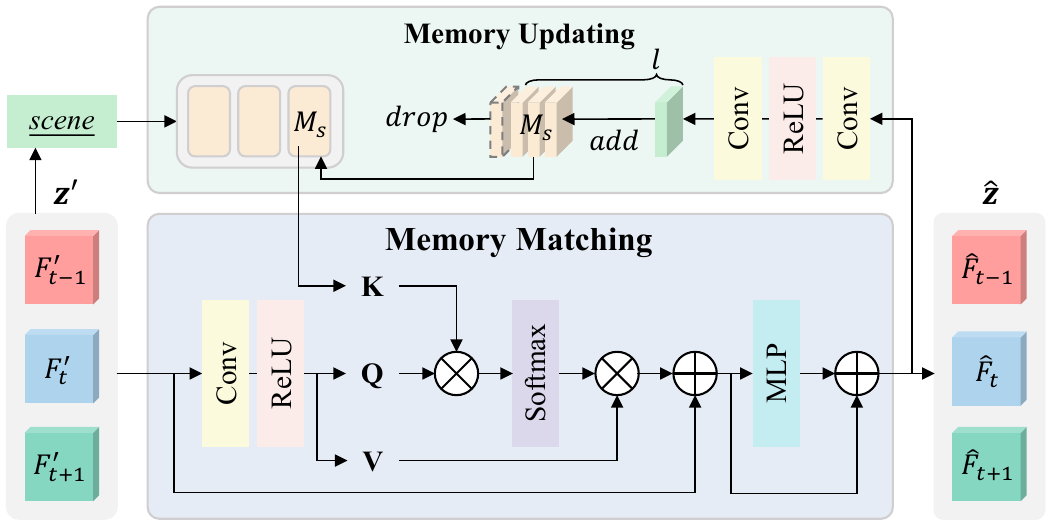}
\end{center}
\caption{DMM processes input features based on the scene and consists of two key components: memory matching and memory updating. In memory matching, the input feature \( F^{\prime}_{t} \) and the retrieved memory feature \( M_{s} \) are used to compute cross-attention, generating the enhanced feature \( \hat{F}_{t} \). Memory updating then refines \( \hat{F}_{t} \) into \( \hat{M}_s \) and adds it to the memory dictionary, ensuring that the dictionary remains both real-time and globally representative.}
\label{fig:memory}
\end{figure}

It is highlighted in blue in Figure~\ref{fig:memory}, takes the processed features from T-MoE, denoted as $\mathbf{z^{\prime}}=\{F^{\prime}_t \in \mathbb{R}^{\frac{H}{2} \times \frac{W}{2} \times C}\}^{T}_{t=1}$, as input for the DMM. It processes video frames sequentially within each scene and dynamically updates its memory dictionary.  
For a given frame at time $t$, its feature representation $F^{\prime}_t$ undergoes dimensionality reduction via a convolutional layer and is then activated with the ReLU function:  
$F^{\prime\prime}_t = \text{ReLU}(\text{Conv}(F^{\prime}_t))$.
This transformation maps the features into a lower-dimensional space, reducing the computational complexity of the memory module.  
Next, the global memory $M_s$ corresponding to the scene is retrieved from the memory dictionary using the scene name $\underline{scene}$ as the key. 
The retrieved memory is mapped to the key matrix $K \in \mathbb{R}^{n \times m \times c}$, while $F^{\prime\prime}_t$ is mapped to the query $Q \in \mathbb{R}^{m \times c}$ and value $V \in \mathbb{R}^{m \times c}$. The $n$, $m$, and $c$ denote the memory unit size, sequence length, and hidden layer dimension, respectively. This process can be formulated as:
\begin{equation}
    K=M_{s}P_{K}, \space Q=F^{\prime\prime}_{t}P_{Q}, \space V=F^{\prime\prime}_{t}P_{V}, 
\end{equation}
where \( P_Q \), \( P_K \), and \( P_V \) are the projection matrices that map the input features into the query, key, and value spaces, respectively. The cross-attention mechanism is then computed using \( Q \), \( K \), and \( V \) as follows:
\begin{equation}
    \begin{aligned}
        &attn=\text{Softmax}(QK^{T})V+F^{\prime}_t,\\
        &\hat{F}_{t}=\text{MLP}(attn)+attn.
    \end{aligned}
\end{equation}

It is important to note that global information is stored and retrieved on a per-scene basis. When processing a completely new scene, no relevant information is available, and the cross-attention mechanism cannot operate effectively. To handle this situation, we introduce a discrimination mechanism: if the input comes from a new scene, we bypass the cross-attention calculation and directly feed the input into an MLP with residual connections for optimization. This design ensures that the memory module can effectively manage inputs from different scenes.

\subsubsection{Memory Updating}
As illustrated in the green part of Figure~\ref{fig:memory}, the memory update module ensures that the memory remains both global and real-time. Specifically, the memory-matching result \( \hat{F}_t \) undergoes two convolutional layers, followed by a ReLU activation, to generate the updated memory feature:  
$\hat{M}_s=\text{Conv}(\text{ReLU}(\text{Conv}(\hat{F}_t)))$.
This updated memory \( \hat{M}_s \) is then used to refine the stored memory. 
To maintain a manageable memory size while preserving its effectiveness, we constrain the memory length for each scene to \( l \) and employ a queue-based update strategy. 
If the memory length is less than \( l \), \( \hat{M}_s \) is appended to the queue; otherwise, the oldest entry is removed before adding \( \hat{M}_s \). This approach prevents excessive memory growth and ensures that stored information remains relevant and up-to-date.  
In our experiments, we set \( l = 2 \). The memory dictionary in DMM is organized by scene, meaning each scene has its own independent memory space, eliminating interference between different videos. During both querying and updating, only the memory associated with the current scene is accessed. 
This design effectively minimizes feature interference across different videos while allowing flexible memory allocation within available computational resources, thereby maximizing the effectiveness of the memory module.

\subsection{Loss Function}
WMNet employs a two-stage training framework. Phase I focuses on improving basic reconstruction ability, while Phase II enhances perceptual quality and structural consistency.

In Phase I, which focuses on self-reconstruction pretraining, we adopt the \( L_1 \) loss to ensure pixel-wise fidelity between the reconstructed and ground-truth videos, which is defined as:  
\begin{equation}
    L_1 = \frac{1}{TN}\sum_{t=1}^{T}\sum_{n=1}^{N}\left | \hat y_{tn}-y_{tn} \right |,
\label{equ:l1}
\end{equation}
where \( \hat{y} \) denotes the reconstructed video, \( y \) represents the ground-truth video, \( t \) indexes the temporal dimension, and \( n \) denotes the pixel position.  

In Phase II, we enhance perceptual quality by introducing the SSIM loss alongside the \( L_1 \) loss. It aims to improve the structural consistency between the reconstructed and target videos and is formulated as:  
\begin{equation}
L_s = 1 - \frac{1}{T}\sum_{T}^{t=1}\text{SSIM}(\hat{y_t},y_t).
\end{equation}

The $\text{SSIM}(\cdot)$ is computed as follows:
\begin{equation}
\text{SSIM}(\hat{y_t},y_t) = \frac{(2\mu_{\hat{y}_t }\mu_{y_t} + C_1)(2\sigma_{\hat{y}_t y_t}+C_2)}{(\mu_{\hat{y}_t}^2+\mu_{y_t}^2+ C_1)(\sigma_{\hat{y}_t}^2+\sigma_{y_t}^2+C_2)},  
\end{equation}
where \(\mu_{\hat{y}_t}\) and \(\mu_{y_t}\) are the mean intensities of the video frames \(\hat{y_t}\) and \(y_t\), respectively, \(\sigma_{\hat{y}_t}\) and \(\sigma_{y_t}\) denote their variances, and \(\sigma_{\hat{y}_t y_t}\) represents their covariance. The constants \(C_1\) and \(C_2\) are included to stabilize the computation and prevent division by zero. 
In summary, the total loss function for Phase II is formulated as a weighted sum of the $L1$ loss and the SSIM loss $L_s$:
\begin{equation}
    L_{total}= L_1+\lambda L_s,
\end{equation}
where $\lambda$ is a hyperparameter that controls the weight of the SSIM loss relative to the $L_1$ loss. In this work, we set \(\lambda = 1\).

\section{Experimental Results}
\label{sec:results}

\subsection{Experimental Setup}

\subsubsection{Dataset}
The dataset used in this experiment comes from HDRTV4K \cite{guo2023learning}, a dataset introduced by Guo et al. in 2023 for HDR video reconstruction.  
HDRTV4K includes HDRTV1K, UGC, and natural scene data. 
However, since it is not organized by scene, we filter and reorganize it to better suit HDR video reconstruction tasks.  
Specifically, we select video frames from the HDRTV4K training set based on scene consistency, ensuring that each scene contains at least eight frames. This process results in 100 scenes with 1,092 video frame pairs.  
Since the HDRTV4K test set is not publicly available, we randomly select 20 scenes from our curated dataset for testing, forming the proposed~\textbf{HDRTV4K-Scene} dataset. The training set consists of 80 scenes with 881 video frame pairs, while the test set contains 20 scenes with 211 video frame pairs.  
To ensure a fair comparison with existing methods, the LDR video data follows the YouTube compression algorithm, consistent with HDRTV1K. Additionally, due to the limited number of consecutive frames per scene in HDRTV4K, we collect new data from 10 scenes following the approach of Chen et al.~\cite{chen2021new}. Each scene contains at least 26 frames, allowing us to assess the model’s ability to reconstruct longer videos. This extended dataset is called~\textbf{HDRTV-LongScene}.
\revise{Furthermore, we adopt the test set of Real-HDRV \cite{shu2024towards} for generalization evaluation. The test set comprises 50 HDR scene videos, each containing 8 frames, and covers a broad range of diverse real-world scenarios.}

\subsubsection{Training Details}
In Phase I, we apply random masks to the low-frequency components of the wavelet transform, gradually increasing the mask rate from 0 to 0.5. The mask size is set to \(1 \times 1\). The encoder consists of 15 residual blocks with a hidden layer size of 64. The initial learning rate is \(2 \times 10^{-4}\). Training runs for 100,000 iterations with a batch size of 16, where each batch contains 8 adjacent frames.  
In Phase II, the memory module has a hidden layer size of 64, and each scene maintains a memory size of 2. The initial learning rate remains \(2 \times 10^{-4}\), and training follows the same 100,000 iterations and batch size settings as in Phase I.  
Both phases involve cropping image patches from high-resolution frames of either \(2160 \times 3840\) or \(1080 \times 1920\). 
To improve data loading efficiency, we pre-crop the training data into \(128 \times 128\) patches with a step size of 240.  
We use the Adam optimizer with decay rates of \(\beta_1 = 0.9\) and \(\beta_2 = 0.99\) for optimization. The learning rate scheduler follows MultiStepLR, reducing the learning rate by half every 20,000 iterations. 
All experiments are conducted on an NVIDIA RTX V100 GPU.

\begin{table*}[t]
\centering
\caption{
Quantitative comparison of various methods on the HDRTV4K-Scene dataset. 
The \colorbox{red!30}{best}, \colorbox{orange!30}{second-best}, and \colorbox{yellow!30}{third-best} results are highlighted for each metric. 
}
\begin{tabular}{l|l|cccccccc}
\toprule
\toprule
Methods&Params& PSNR$\uparrow$ & SSIM$\uparrow$ & SR-SIM $\uparrow$ & $\Delta E_{ITP}$  $\downarrow$& HDR-VDP3 $\uparrow$& LPIPS$\downarrow$ & $E_{warp}$$\downarrow$\\ 
\midrule
HuoPhyEO~\cite{huo2014physiological} &-& 23.60 & 0.9012 & 0.9847 & 48.55 & 6.618 & 0.1411 & 0.01221\\
KovaleskiEO~\cite{kovaleski2014high} &-& 26.92 & 0.9187 & 0.9938 & 30.88 & 7.014 & 0.1309 & 0.01035\\
\midrule
Ada-3DLUT~\cite{zeng2020learning} & 594K & 28.20 & 0.9202 & 0.9919 & 27.84 & 7.195 & 0.1131 & 0.01869\\
CSRNet~\cite{he2020conditional} & 36K & 33.12 & 0.9424 & 0.9953 & 15.48 & 7.553 & 0.0920 & 0.01417 \\
HDRTVNet-w/o HG~\cite{chen2021new} & 1.41M & \cellcolor{yellow!30}35.45 & 0.9577 & \cellcolor{yellow!30}0.9981 & 11.86 & 8.008 & 0.0735 & 0.01404 \\
HDRTVNet~\cite{chen2021new} & 37.2M & 35.41 & 0.9576 & 0.9975 & 11.86 & 8.007 & 0.0736 & \cellcolor{orange!30}0.01394 \\
FMNet~\cite{xu2022fmnet} & 1.24M & \cellcolor{orange!30}35.51 & \cellcolor{orange!30}0.9592 & \cellcolor{orange!30}0.9983 & \cellcolor{orange!30}11.46 & 7.920 & 0.0747 & 0.01448 \\
HDRTVNet++-w/o HG~\cite{chen2023towards} & 591K & 35.38 & \cellcolor{yellow!30}0.9580 & 0.9978 & \cellcolor{yellow!30}11.83 & \cellcolor{red!30}8.200 & \cellcolor{yellow!30}0.0724 & \cellcolor{yellow!30}0.01403 \\
HDRTVNet++~\cite{chen2023towards} & 4.98M & 34.77 & 0.9526 & 0.9976 & 13.33 & \cellcolor{yellow!30}8.162 & \cellcolor{orange!30}0.0657 & 0.01419 \\
Ours & 1.36M &  \cellcolor{red!30}{36.23} & \cellcolor{red!30}{0.9630} & \cellcolor{red!30}{0.9987} & \cellcolor{red!30}{10.78} & \cellcolor{orange!30}{8.173} &\cellcolor{red!30}{0.0641}&\cellcolor{red!30}{0.01392} \\ 	 	 	
\bottomrule
\bottomrule
\end{tabular}
\label{tab:scene}
\end{table*}

\begin{table*}[t]
\centering
\caption{
Quantitative comparison of various methods on the HDRTV4K-LongScene dataset.  
The \colorbox{red!30}{best}, \colorbox{orange!30}{second-best}, and \colorbox{yellow!30}{third-best} results are highlighted for each metric.
}
\begin{tabular}{l|l|ccccccc}
\toprule
\toprule
Methods&Params& PSNR$\uparrow$ & SSIM$\uparrow$ & SR-SIM $\uparrow$ & $\Delta E_{ITP}$  $\downarrow$& HDR-VDP3 $\uparrow$& LPIPS$\downarrow$ & $E_{warp}$$\downarrow$\\ 
\midrule
HuoPhyEO~\cite{huo2014physiological} &-& 26.88 & 0.9347 & 0.9945 & 38.43 & 7.620 & 0.0962 & 0.00171\\
KovaleskiEO~\cite{kovaleski2014high} &-& 27.46 & 0.9294 & 0.9951 & 32.88 & 7.596 & 0.1019 & 0.00158\\
\midrule
Ada-3DLUT~\cite{zeng2020learning} & 594K & 27.59 & 0.9357 & 0.9964 & 36.84 & 7.796 & 0.0957 & 0.00253\\
CSRNet~\cite{he2020conditional} & 36K & \cellcolor{orange!30}37.09 & 0.9754 & 0.9981 & 15.09 & 8.401 & 0.0595 & 0.00215\\
HDRTVNet-w/o HG~\cite{chen2021new} & 1.41M & 36.61 & 0.9776 & 0.9983 & 14.79 & 8.464 & \cellcolor{red!30}0.0590 & 0.00204\\
HDRTVNet~\cite{chen2021new} & 37.2M & 36.61 & 0.9776 & 0.9983 & 14.79 & \cellcolor{yellow!30}8.490 & \cellcolor{orange!30}0.0591 & 0.00204\\
FMNet~\cite{xu2022fmnet} & 1.24M & \cellcolor{yellow!30}37.05 & \cellcolor{orange!30}0.9789 & \cellcolor{yellow!30}0.9984 & \cellcolor{yellow!30}14.66 & 8.464 & 0.0621 & \cellcolor{red!30}0.00183\\
HDRTVNet++-w/o HG~\cite{chen2023towards} & 591K & 36.78 & \cellcolor{yellow!30}0.9777 & 0.9986 & \cellcolor{orange!30}14.64 & \cellcolor{orange!30}8.508 & \cellcolor{yellow!30}0.0597 & 0.00210\\
HDRTVNet++~\cite{chen2023towards} & 4.98M & 35.61 & 0.9659 & \cellcolor{orange!30}0.9985 & 16.85 & 8.443 & 0.0600 & \cellcolor{yellow!30}0.00187\\
Ours & 1.36M &  \cellcolor{red!30}{37.19} & \cellcolor{red!30}{0.9803} & \cellcolor{red!30}{0.9986} & \cellcolor{red!30}{14.43} & \cellcolor{red!30}{8.591} &\cellcolor{orange!30}{0.0591}&\cellcolor{orange!30}{0.00186}\\  	
\bottomrule
\bottomrule
\end{tabular}
\label{tab:longscene}
\end{table*}

\subsubsection{Evaluation Metrics}
Following previous work~\cite{chen2023towards, xu2022fmnet}, we employ seven evaluation metrics for a comprehensive comparison: PSNR, SSIM, SR-SIM~\cite{zhang2012sr}, HDR-VDP3 \cite{mantiuk2011hdr}, $\Delta E_{ITP}$, LPIPS~\cite{zhang2018unreasonable}, and $E_{warp}$~\cite{teed2020raft}. SSIM and SR-SIM are widely used for measuring image similarity. To assess color differences specific to HDRTV, we use $\Delta E_{ITP}$, which quantifies color deviations in real HDRTV content, where lower values indicate better color matching.  
HDR-VDP3, an improved version of HDR-VDP2, supports the Rec.2020 color space and evaluates signal similarity in linear light space based on the human visual system, with higher scores indicating less visual degradation. For HDR-VDP3, we configure the comparison as ``side-by-side," set the color encoding to ``rgb-bt.2020," use a pixel density of 50 pixels per degree, and enable the ``led-lcd-wcg" option under ``rgb-display."  
To better align with human perception, we use LPIPS to measure perceptual differences between HDR images. Since our experiments focus on videos, we also introduce $E_{warp}$ to assess temporal consistency between frames.

\begin{figure*}[t]
\begin{center}
\includegraphics[width=1.0\linewidth]{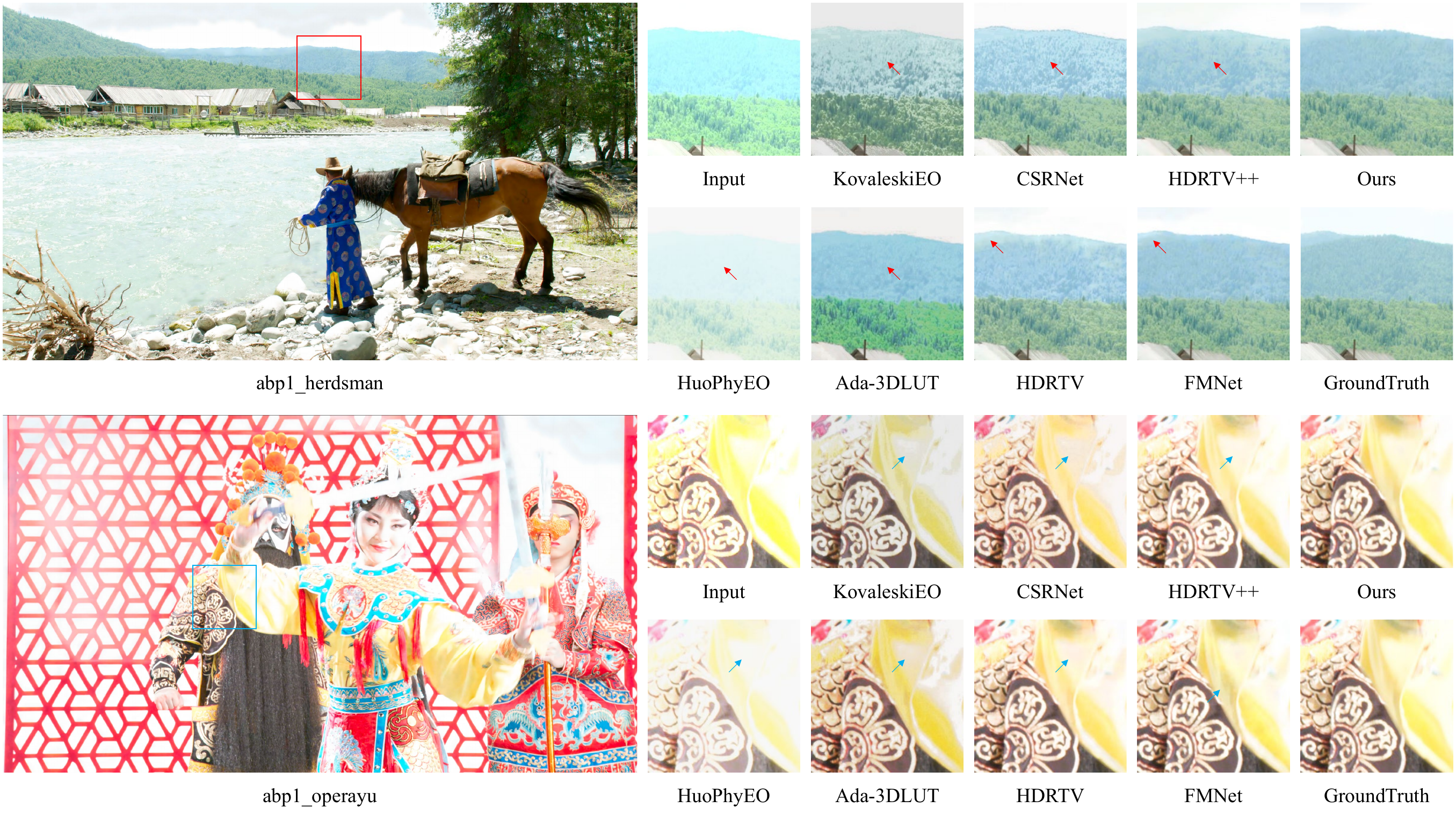}
\end{center}
\caption{Qualitative results on the HDRTV4K-Scene dataset, please zoom in for a better view of details. }
\label{fig:qualitative}
\end{figure*}

\subsection{Experimental Comparison} 

\subsubsection{Quantitative Comparison} 
To evaluate the performance of the proposed WMNet, we compare it with both traditional and state-of-the-art methods, including HuoPhyEO~\cite{huo2014physiological}, KovaleskiEO~\cite{kovaleski2014high}, Ada-3DLUT~\cite{zeng2020learning}, CSRNet~\cite{he2020conditional}, HDRTVNet~\cite{chen2021new}, FMNet~\cite{xu2022fmnet}, and HDRTVNet++~\cite{chen2023towards}.  
Following Chen et al.~\cite{chen2023towards}, we also evaluate HDRTVNet and HDRTVNet++ without highlight generation (HG).  
The previous methods are also compared with Deep SR-ITM~\cite{kim2019deepsr} and JSI-GAN~\cite{kim2020jsi}, but these two methods are used for joint super-resolution and HDR video reconstruction tasks, which is unfair to compare with methods that only perform HDR video reconstruction. Therefore, these two methods are not compared in this paper.

The quantitative results on the HDRTV4K-Scene dataset are presented in Table~\ref{tab:scene}. 
Compared to existing deep learning-based methods, WMNet achieves optimal or near-optimal performance across all metrics, demonstrating its effectiveness.  
In terms of PSNR and SSIM, WMNet surpasses the second-best method, FMNet, by 0.72 dB and 0.0038, respectively, indicating superior reconstruction quality. 
It also achieves the best performance on SR-SIM and reduces the \(\Delta E_{ITP}\) score to 10.78, 0.68 lower than the second-best method, highlighting its advantage in color reconstruction accuracy. Furthermore, WMNet performs favorably on HDR-VDP3 and LPIPS, while excelling in \(E_{warp}\), demonstrating strong temporal consistency.  
It is worth noting that traditional methods exhibit better temporal consistency than deep learning–based approaches, as they apply uniform processing to all video frames, preserving temporal coherence but at the expense of reconstruction quality.  
Table~\ref{tab:longscene} presents the results on the HDRTV4K-LongScene test set, further validating WMNet’s effectiveness on extended video sequences. 
Our method achieves the best or second-best results across all metrics, demonstrating its capability to handle long video content. 
In summary, the results in Tables~\ref{tab:scene} and \ref{tab:longscene} confirm the effectiveness of the proposed WMNet for HDR video reconstruction.

\begin{table*}[t]
\centering
\caption{
\revise{Quantitative generalization comparison of various methods on the RealHDRV dataset.  
The \colorbox{red!30}{best}, \colorbox{orange!30}{second-best}, and \colorbox{yellow!30}{third-best} results are highlighted for each metric.
The inference time is measured on a batch of eight 128 $\times$ 128 frames and reported as the average time per frame.}
}
\begin{tabular}{l|l|cccccccc}
\toprule
\toprule
Methods&Params& PSNR$\uparrow$ & SSIM$\uparrow$ & SR-SIM $\uparrow$ & $\Delta E_{ITP}$  $\downarrow$& HDR-VDP3 $\uparrow$& LPIPS$\downarrow$ & time(ms)$\downarrow$\\ 
\midrule
HuoPhyEO~\cite{huo2014physiological} &-& 26.67 & 0.9057 & 0.9780 & 31.90 & \cellcolor{red!30}7.591 & 0.1114 & -\\
KovaleskiEO~\cite{kovaleski2014high} &-& 22.04 & 0.8214 & 0.9659 & 56.17 & 6.263 & 0.1593 & -\\
\midrule
Ada-3DLUT~\cite{zeng2020learning} & 594K & 22.88 & 0.8470 & \cellcolor{red!30}0.9825 & 53.42 & 6.983 & 0.1583 & \cellcolor{red!30}0.5772\\
CSRNet~\cite{he2020conditional} & 36K & 27.28 & 0.9220 & 0.9793 & 31.05 & 6.985 & 0.1118 & \cellcolor{yellow!30}1.0456\\
HDRTVNet-w/o HG~\cite{chen2021new} & 1.41M & 27.21 & \cellcolor{yellow!30}0.9506 & 0.9802 & \cellcolor{yellow!30}29.71 & 7.112 & \cellcolor{orange!30}0.0805 & 3.7391\\
HDRTVNet~\cite{chen2021new} & 37.2M & 27.22 & \cellcolor{yellow!30}0.9506 & 0.9803 & 29.75 & 7.114 & \cellcolor{yellow!30}0.0804 & 6.5009\\
FMNet~\cite{xu2022fmnet} & 1.24M & \cellcolor{red!30}27.39 & 0.9505 & 0.9801 & \cellcolor{red!30}29.32 & \cellcolor{yellow!30}7.161 & 0.0809 & 2.3932\\
HDRTVNet++-w/o HG~\cite{chen2023towards} & 591K & 27.24 & 0.9371 & 0.9805 & 30.73 & 7.085 & 0.0837 & 12.698\\
HDRTVNet++~\cite{chen2023towards} & 4.98M & \cellcolor{yellow!30}27.32 & 0.9361 & \cellcolor{orange!30}0.9821 & 30.13 & 7.132 & 0.0889 & 20.108\\
Ours & 1.36M &  \cellcolor{red!30}27.39 & \cellcolor{red!30}0.9507 & \cellcolor{yellow!30}0.9807 & \cellcolor{orange!30}29.47 & \cellcolor{orange!30}7.200 & \cellcolor{red!30}0.0800 & \cellcolor{orange!30}0.9355\\  	
\bottomrule
\bottomrule
\end{tabular}
\label{tab:generalization}
\end{table*}

\begin{figure*}[t]
\begin{center}
\includegraphics[width=1.0\linewidth]{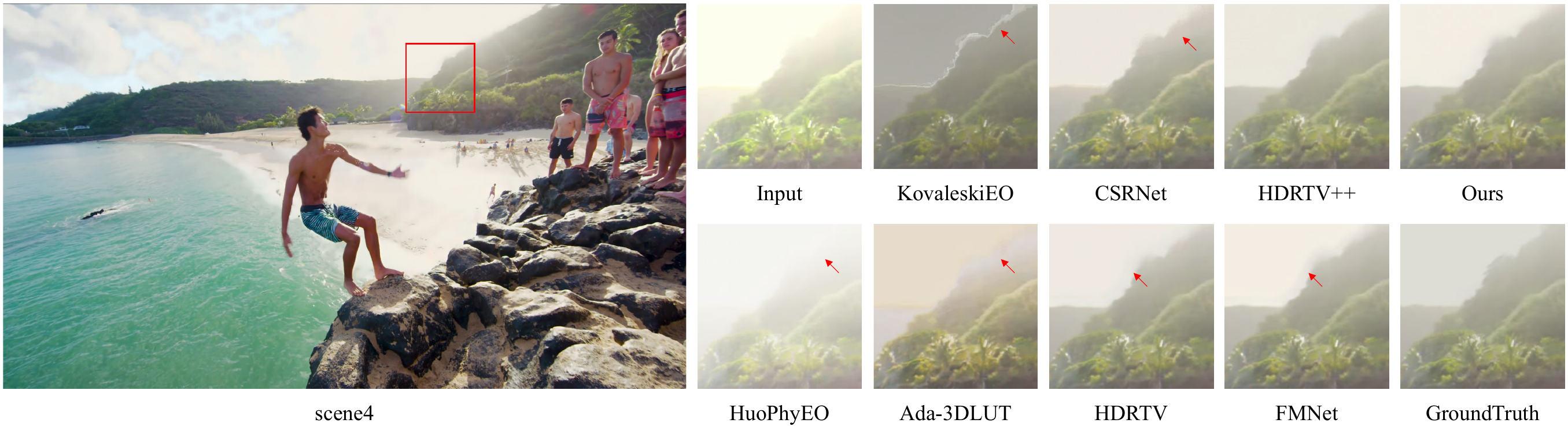}
\end{center}
\caption{Qualitative results on the HDRTV4K-LongScene dataset, please zoom in for a better view of details.}
\label{fig:qualitative2}
\end{figure*}

\subsubsection{Qualitative Comparison}
We compare the qualitative results of WMNet with other methods. Since HDR video frames cannot be directly displayed on standard dynamic range (SDR) screens, we apply tone mapping to adapt them for SDR viewing. 
To ensure a more accurate display, we follow the approach in Deep-SR-ITM~\cite{kim2019deepsr} and use the MADVR renderer to tone map video frames for comparative analysis.  
Figure~\ref{fig:qualitative} presents qualitative results on HDRTV4K-Scene. 
Traditional methods, such as HuoPhyEO and KovaleskiEO, struggle to accurately restore brightness and contrast due to their reliance on manually set parameters. 
Ada-3DLUT and CSRNet exhibit noticeable color discrepancies, indicating limitations in color restoration. 
HDRTV++ fails to preserve fine details, such as textures on mountains and clothing patterns, while HDRTV and FMNet introduce visible artifacts, degrading visual quality. 
In contrast, WMNet effectively restores color and detail, producing results that are visually closer to the ground truth.  
Figure~\ref{fig:qualitative2} shows qualitative results on HDRTV4K-LongScene, where WMNet also achieves the best performance in color fidelity and detail preservation compared to other methods.

\subsubsection{\revise{Generalization Comparison}}
\revise{
We further evaluate the generalization capability of WMNet on the cross-domain Real-HDRV dataset, which contains diverse lighting conditions, camera motions, and color variations. 
As shown in Table~\ref{tab:generalization}, WMNet consistently ranks among the top three across all evaluation metrics, clearly demonstrating its robustness and strong generalization ability beyond the training domain.
In addition to reconstruction quality, we also assess inference efficiency. 
Benefiting from a lightweight architecture and efficient temporal modeling through T-MoE, WMNet achieves the second-fastest inference speed of 0.9355 ms per frame, outperforming most competing methods, including computation-heavy architectures such as FMNet and HDRTVNet++. 
This result indicates that high reconstruction performance can be achieved without incurring excessive computational cost, validating the overall efficiency of our model design.
}

\begin{figure}[t]
\begin{center}
\includegraphics[width=1.0\linewidth]{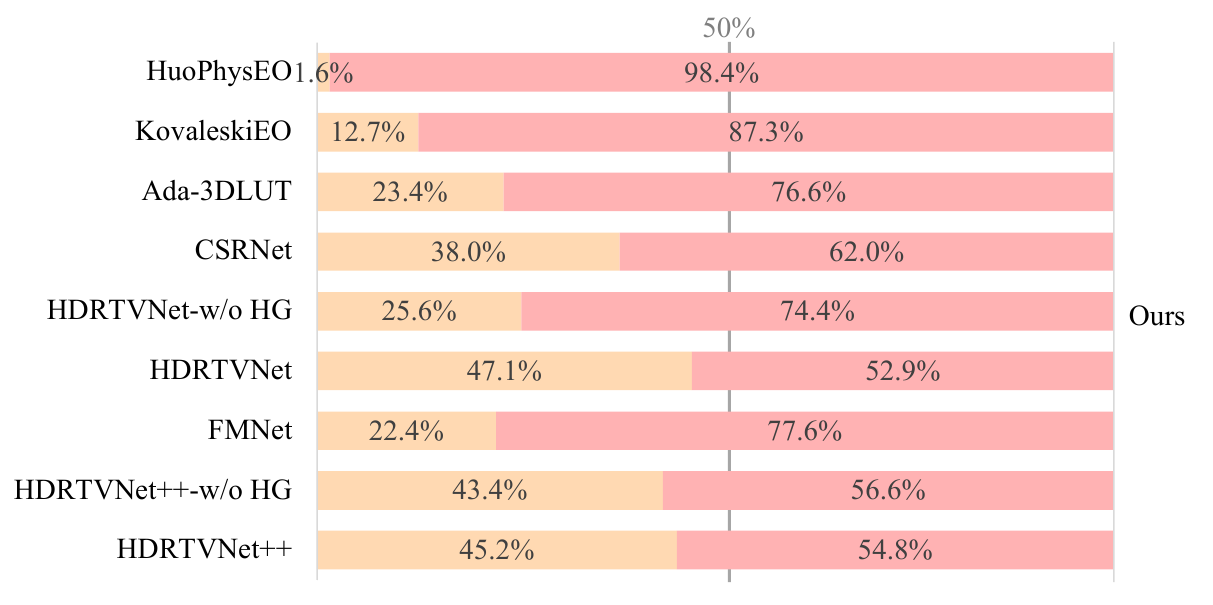}
\end{center}
\caption{\revise{User Study Statistics. The \colorbox{red!30}{red} portion represents the percentage of users who rated our method's output as better, while the \colorbox{orange!30}{orange} portion represents the percentage who preferred the output of the compared method.}}
\label{fig:user_study}
\end{figure}

\subsubsection{\revise{User Study}}
\revise{
We conduct a user study to provide comprehensive subjective evidence for evaluating the visual performance of our method. 
Participants compare our results with those generated by state-of-the-art approaches in a pairwise preference test under controlled viewing conditions.
As shown in Figure~\ref{fig:user_study}, our method consistently receives higher preference rates than all competing methods. 
In each bar, the red region indicates the proportion of users who prefer our results, while the orange region corresponds to the compared method. 
Our method achieves an overwhelming preference rate of 98.4\% over HuoPhysEO1. 
Even when compared with strong deep learning baselines such as FMNet (77.6\%) and CSRNet (62.0\%), our method remains clearly favored.
In addition, our method maintains a stable advantage over HDR-TVNet variants, with preference rates ranging from 52.9\% to 74.4\%. 
These findings demonstrate that our method not only improves objective reconstruction metrics but also aligns more closely with human visual perception, producing more natural, vivid, and visually appealing HDR video results.
}

\begin{table}[t]
\centering
\caption{Ablation experiments on the proposed modules conducted on the HDRTV4K-Scene dataset.}
\setlength{\tabcolsep}{2.5pt}
\begin{tabular}{cccc|c|ccc}
\toprule
\toprule
Baseline & W-MIM & T-MoE & DMM & Param. & PSNR $\uparrow$ & SSIM $\uparrow$  & $\Delta E_{ITP}$$\downarrow$\\ 
\midrule
 \checkmark & &  &  & 1.30M & 35.15 & 0.9612 & 12.11\\
 \checkmark & \checkmark &  &  & 1.30M & \cellcolor{yellow!30}{36.00} &\cellcolor{yellow!30}{0.9627} & \cellcolor{yellow!30}11.05\\
 \checkmark & \checkmark & \checkmark &  & 1.35M & \cellcolor{orange!30}{36.17}& \cellcolor{orange!30}{0.9629} & \cellcolor{orange!30}10.85\\
 \checkmark & \checkmark & \checkmark & \checkmark & 
1.36M &\cellcolor{red!30}{36.23} & \cellcolor{red!30}{0.9630} &  \cellcolor{red!30}10.78\\	 	 	
\bottomrule
\bottomrule
\end{tabular}
\label{tab:modules}
\end{table}

\subsection{Ablation Studies}
In this subsection, we conduct a series of ablation studies to comprehensively evaluate the contribution of each component in WMNet and the impact of different hyperparameter choices. All training and testing are performed on the HDRTV4K-Scene dataset to ensure consistency and reliability in our analysis.

\subsubsection{Effectiveness of the W-MIM} 
We validate the effectiveness of W-MIM pre-training by comparing results with and without this step, as shown in the first and second rows of Table~\ref{tab:modules}, where W-MIM significantly enhances HDR video reconstruction by guiding the model to focus on color fidelity and fine details. 
Additionally, we conduct ablation studies on the W-MIM masking scheme by evaluating wavelet domain masked image modeling in low-frequency (W-MIM-L) and high-frequency components (W-MIM-H), as well as spatial domain masked image modeling (S-MIM). 
W-MIM integrates masking strategies across both frequency domains, improving the model’s ability to recover color and fine-grained details. 
Table~\ref{tab:mask} presents the results of individually applying W-MIM-L, W-MIM-H, and S-MIM, all of which contribute to enhanced reconstruction quality. 
Further experiments show that our W-MIM, which combines W-MIM-L and W-MIM-H, achieves the best performance, particularly in color accuracy, reinforcing the motivation behind our proposed W-MIM approach.

\begin{table}[t]
\centering
\caption{\revise{Ablation experiments on wavelet domain masking.}}
\setlength{\tabcolsep}{6pt}
\begin{tabular}{ccc|cccc}
\toprule
\toprule
S-MIM & W-MIM-L & W-MIM-H & PSNR$\uparrow$ & SSIM $\uparrow$  & $\Delta E_{ITP}$$\downarrow$\\ 
\midrule
  &  &  & 36.07 & 0.9626 & 11.01 \\
\checkmark &   &  & {36.12} & {0.9627}  &{10.97}  \\
& \checkmark  & & \cellcolor{orange!30}36.19 & \cellcolor{orange!30}0.9629 & \cellcolor{yellow!30}10.87  \\	
 & & \checkmark  & \cellcolor{yellow!30}{36.17} &\cellcolor{orange!30}{0.9629} &\cellcolor{red!30}{10.77}   \\	
 & \checkmark & \checkmark  & \cellcolor{red!30}{36.23}& \cellcolor{red!30}{0.9630} & \cellcolor{orange!30}{10.78}  \\	
\bottomrule
\bottomrule
\end{tabular}
\label{tab:mask}
\end{table}

\subsubsection{Effectiveness of the T-MoE}
We evaluate the impact of the T-MoE module, which enhances HDR video reconstruction by effectively leveraging information from adjacent frames to mitigate temporal inconsistencies and improve overall coherence. 
By dynamically selecting the most relevant features from neighboring frames, T-MoE refines the reconstruction process, reducing flickering artifacts and enhancing motion continuity. 
As evidenced by the second and third rows of Table~\ref{tab:modules}, incorporating T-MoE leads to notable improvements in both perceptual and quantitative metrics, demonstrating its effectiveness in capturing temporal dependencies and refining fine details across consecutive frames.

\subsubsection{Effectiveness of the DMM} 
\revise{We examine the role of the DMM module, where the comparison between the third and fourth rows of Table~\ref{tab:modules} demonstrates that DMM provides a measurable boost in HDR video reconstruction performance by introducing a global memory mechanism.
By maintaining a persistent memory unit per scene, DMM lifts strict temporal access constraints and enables the model to leverage a broader temporal context during reconstruction. This allows the network to retrieve and integrate scene-level information across frames, leading to stronger temporal consistency and more faithful detail recovery.
However, the introduction of this memory mechanism also increases computational overhead, necessitating a careful balance between performance and efficiency. 
To evaluate the trade-off, we conduct an ablation experiment on the size of the memory unit. 
As shown in Table~\ref{tab:mem_size}, increasing the memory unit size leads to a rise in FLOPs while only yielding marginal improvements in reconstruction quality. 
Based on these results, we set the memory unit size to 2, achieving an optimal balance between computational cost and HDR reconstruction performance.}

\begin{table}[t]
\centering
\caption{Ablation experiments on memory unit size.}
\setlength{\tabcolsep}{7.pt}
\begin{tabular}{c|c|cccc}
\toprule
\toprule
Size  & FLOPs (G)  & PSNR$\uparrow$ & SSIM$\uparrow$ & $\Delta E_{ITP}$$\downarrow$ & $E_{warp}$$\downarrow$\\ 
\midrule
1  & 48.48&  \cellcolor{yellow!30}{36.20}  &    \cellcolor{orange!30}{0.9629}   & \cellcolor{yellow!30}{10.81}  & \cellcolor{yellow!30} 0.01558\\	 
2  & 48.51&  \cellcolor{orange!30}36.23  &   \cellcolor{red!30}{0.9630}   & \cellcolor{orange!30}{10.78} &  \cellcolor{red!30}0.01392 \\	 
4  & 48.55& \cellcolor{red!30}{36.24}   &   \cellcolor{red!30}{0.9630}   & \cellcolor{red!30}{10.75}  &  \cellcolor{orange!30}0.01458\\	 
\bottomrule
\bottomrule
\end{tabular}
\label{tab:mem_size}
\end{table}

\begin{table}[t]
\centering
\caption{Ablation experiments on loss function.}
\begin{tabular}{cc|ccccc}
\toprule
\toprule
$L_1$ & $L_s$ & PSNR$\uparrow$ & SSIM $\uparrow$ & SR-SIM $\uparrow$ & $\Delta E_{ITP}$$\downarrow$ & LPIPS$\downarrow$\\ 
\midrule
\checkmark &   & \cellcolor{orange!30}{35.89} &\cellcolor{orange!30}{0.9596} & \cellcolor{orange!30}0.9986 &\cellcolor{orange!30}{11.04}  & \cellcolor{orange!30}0.0738\\
\checkmark & \checkmark  & \cellcolor{red!30}{36.23}& \cellcolor{red!30}{0.9630}& \cellcolor{red!30}0.9987 & \cellcolor{red!30}{10.78} & \cellcolor{red!30}0.0641 \\	
\bottomrule
\bottomrule
\end{tabular}
\label{tab:loss}
\end{table}

\begin{table}[t]
\centering
\caption{\revise{Ablation experiments on wavelet types.}}
\begin{tabular}{c|ccc}
\toprule
\toprule
Type & PSNR$\uparrow$ & SSIM$\uparrow$ & $\Delta E_{ITP}$$\downarrow$\\ 
\midrule
bior4.4 &  \cellcolor{yellow!30}{36.13}  &    \cellcolor{orange!30}{0.9627}   & \cellcolor{orange!30}{10.85} \\	 
sym4 &  \cellcolor{yellow!30}{36.13}  &   \cellcolor{orange!30}{0.9627}   & \cellcolor{yellow!30}{10.86} \\	 
db2 & \cellcolor{orange!30}{36.16}   &   \cellcolor{orange!30}{0.9627}   & \cellcolor{red!30}{10.78} \\	 
Haar & \cellcolor{red!30}{36.23}& \cellcolor{red!30}{0.9630} & \cellcolor{red!30}{10.78} \\	 
\bottomrule
\bottomrule
\end{tabular}
\label{tab:wavelet_type}
\end{table}

\subsubsection{Analysis on Losses}
Most video reconstruction methods primarily rely on the L1 loss function during training, which ensures pixel-wise accuracy but may overlook structural information and fine details. In Phase II of our approach, we introduce the SSIM loss function to guide the network toward better structural awareness, enhancing the preservation of intricate details and improving perceptual quality. As shown in Table~\ref{tab:loss}, the comparison between using only the L1 loss and the combination of L1 and SSIM loss functions demonstrates that incorporating SSIM effectively boosts reconstruction quality. This highlights the importance of structural similarity in refining HDR video reconstruction, leading to more visually coherent and detailed outputs.

\subsubsection{\revise{Ablation of Wavelet Types}}
\revise{We further conduct an ablation study on the choice of wavelet basis, evaluating several commonly used wavelet types, including Haar, bior4.4, sym4, and db2. 
As shown in Table~\ref{tab:wavelet_type}, the performance across different wavelets remains broadly comparable, with only slight variations in reconstruction accuracy and color fidelity. 
Notably, Haar achieves the best overall results. 
Considering its compact support, orthogonality, and low computational overhead, Haar offers a favorable balance between efficiency and reconstruction quality. 
Therefore, we adopt Haar as the default wavelet basis in WMNet.}

\subsubsection{\revise{Impact of Decomposition Count}}
\revise{We evaluate the effect of varying the number of wavelet decomposition levels from 1 to 3. 
As shown in Table~\ref{tab:decom_count}, increasing the decomposition depth leads to consistent performance gains. 
In particular, using three decomposition levels achieves the highest PSNR and SSIM, along with a lower \(\Delta E_{ITP}\), indicating improved structural fidelity and color accuracy. 
These results verify that deeper decomposition facilitates more effective multi-scale feature representation in the wavelet domain. Although one-level decomposition yields acceptable performance, the improvements observed at three levels justify its selection. 
Therefore, we adopt three-level decomposition as the default setting in WMNet.}

\begin{table}[t]
\centering
\caption{\revise{Ablation experiments on decomposition count.}}
\begin{tabular}{c|ccc}
\toprule
\toprule
Type & PSNR$\uparrow$ & SSIM$\uparrow$ & $\Delta E_{ITP}$$\downarrow$\\ 
\midrule
1 &  \cellcolor{orange!30}{36.19}  &    \cellcolor{orange!30}{0.9629}   & \cellcolor{yellow!30}{10.81} \\	 
2 &  \cellcolor{yellow!30}{36.17}  &   \cellcolor{orange!30}{0.9629}   & \cellcolor{red!30}{10.77} \\	 
3 & \cellcolor{red!30}{36.23}& \cellcolor{red!30}{0.9630} & \cellcolor{orange!30}{10.78} \\	 
\bottomrule
\bottomrule
\end{tabular}
\label{tab:decom_count}
\end{table}

\section {Conclusions} 
\label{sec:conclusion}
\revise{
This paper presents WMNet, a novel HDR video reconstruction framework designed to address the challenges of color inaccuracy and temporal inconsistency in existing methods. 
WMNet employs a two-phase training strategy that integrates Wavelet-domain Masked Image Modeling (W-MIM) with temporal modeling. 
In Phase I, W-MIM performs self-reconstruction pre-training by selectively masking color and detail components in the wavelet domain, enabling the network to learn robust and generalizable color restoration. 
In Phase II, the model is fine-tuned to improve overall reconstruction fidelity.
To improve temporal coherence, WMNet introduces a Temporal Mixture of Experts (T-MoE) module for adaptive frame fusion and a Dynamic Memory Module (DMM) to capture long-range dependencies.
We also reorganize the HDRTV4K dataset into HDRTV4K-LongScene, providing a reliable benchmark for scene-based HDR video reconstruction. 
Extensive experiments demonstrate that WMNet achieves state-of-the-art performance across multiple metrics, and cross-domain evaluation on Real-HDRV further verifies its robustness and generalization capability.
}

\bibliography{ref}

@inproceedings{kim2019deepsr,
  title={{Deep SR-ITM}: Joint learning of super-resolution and inverse tone-mapping for 4k {UHD HDR} applications},
  author={Kim, Soo Ye and Oh, Jihyong and Kim, Munchurl},
  booktitle={Proceedings of the IEEE/CVF International Conference on Computer Vision},
  address = {Seoul, South Korea},
  pages={3116--3125},
  year={2019}
}

@inproceedings{kim2020jsi,
  title={{JSI-GAN}: Gan-based joint super-resolution and inverse tone-mapping with pixel-wise task-specific filters for {UHD HDR} video},
  author={Kim, Soo Ye and Oh, Jihyong and Kim, Munchurl},
  booktitle={Proceedings of the AAAI Conference on Artificial Intelligence},
  volume={34},
  number={07},
  pages={11287--11295},
  address = {New York, NY, USA},
  year={2020}
}

@inproceedings{xu2022fmnet,
  title={{FMNet}: Frequency-aware modulation network for {SDR-to-HDR} translation},
  author={Xu, Gang and Hou, Qibin and Zhang, Le and Cheng, Ming-Ming},
  booktitle={Proceedings of the 30th ACM International Conference on Multimedia},
  pages={6425--6435},
  year={2022},
  address = {Lisbon, Portugal}
}

@inproceedings{chen2021new,
  title={A new journey from {SDRTV} to {HDRTV}},
  author={Chen, Xiangyu and Zhang, Zhengwen and Ren, Jimmy S and Tian, Lynhoo and Qiao, Yu and Dong, Chao},
  booktitle={Proceedings of the IEEE/CVF International Conference on Computer Vision},
  pages={4500--4509},
  address = {Montreal, QC, Canada},
  year={2021}
}

@article{chen2023towards,
  title={Towards Efficient {SDRTV}-to-{HDRTV} by Learning from Image Formation},
  author={Chen, Xiangyu and Li, Zheyuan and Zhang, Zhengwen and Ren, Jimmy S and Liu, Yihao and He, Jingwen and Qiao, Yu and Zhou, Jiantao and Dong, Chao},
  journal={arXiv preprint arXiv:2309.04084},
  year={2023}
}

@inproceedings{he2022masked,
  title={Masked autoencoders are scalable vision learners},
  author={He, Kaiming and Chen, Xinlei and Xie, Saining and Li, Yanghao and Doll{\'a}r, Piotr and Girshick, Ross},
  booktitle={Proceedings of the IEEE/CVF Conference on Computer Vision and Pattern Recognition},
  pages={16000--16009},
  address = {New Orleans, LA, USA},
  year={2022}
}

@inproceedings{xie2022simmim,
  title={Simmim: A simple framework for masked image modeling},
  author={Xie, Zhenda and Zhang, Zheng and Cao, Yue and Lin, Yutong and Bao, Jianmin and Yao, Zhuliang and Dai, Qi and Hu, Han},
  booktitle={Proceedings of the IEEE/CVF Conference on Computer Vision and Pattern Recognition},
  pages={9653--9663},
  year={2022},
  address = {New Orleans, LA, USA}
}

@inproceedings{oh2019video,
  title={Video object segmentation using space-time memory networks},
  author={Oh, Seoung Wug and Lee, Joon-Young and Xu, Ning and Kim, Seon Joo},
  booktitle={Proceedings of the IEEE/CVF International Conference on Computer Vision},
  pages={9226--9235},
  year={2019},
  address = {Long Beach, CA, USA}
}

@inproceedings{kim2019deep,
  title={Deep video inpainting},
  author={Kim, Dahun and Woo, Sanghyun and Lee, Joon-Young and Kweon, In So},
  booktitle={Proceedings of the IEEE/CVF Conference on Computer Vision and Pattern Recognition},
  pages={5792--5801},
  address = {Long Beach, CA, USA},
  year={2019}
}

@inproceedings{ji2022multi,
  title={Multi-scale memory-based video deblurring},
  author={Ji, Bo and Yao, Angela},
  booktitle={Proceedings of the IEEE/CVF Conference on Computer Vision and Pattern Recognition},
  pages={1919--1928},
  address = {New Orleans, LA, USA},
  publisher={IEEE},
  year={2022}
}

@inproceedings{kim2022pin,
  title={Pin the memory: Learning to generalize semantic segmentation},
  author={Kim, Jin and Lee, Jiyoung and Park, Jungin and Min, Dongbo and Sohn, Kwanghoon},
  booktitle={Proceedings of the IEEE/CVF Conference on Computer Vision and Pattern Recognition},
  pages={4350--4360},
  address = {New Orleans, LA, USA},
  publisher={IEEE},
  year={2022}
}

@inproceedings{gao2023memotr,
  title={MeMOTR: Long-Term Memory-Augmented Transformer for Multi-Object Tracking},
  author={Gao, Ruopeng and Wang, Limin},
  booktitle={Proceedings of the IEEE/CVF International Conference on Computer Vision},
  pages={9901--9910},
  address= {Paris, France},
  publisher = {IEEE},
  year={2023}
}

@article{bao2021beit,
  title={Beit: Bert pre-training of image transformers},
  author={Bao, Hangbo and Dong, Li and Piao, Songhao and Wei, Furu},
  journal={arXiv preprint arXiv:2106.08254},
  year={2021}
}

@inproceedings{yoo2019photorealistic,
  title={Photorealistic style transfer via wavelet transforms},
  author={Yoo, Jaejun and Uh, Youngjung and Chun, Sanghyuk and Kang, Byeongkyu and Ha, Jung-Woo},
  booktitle={Proceedings of the IEEE/CVF International Conference on Computer Vision},
  pages={9036--9045},
  year={2019},
  address = {Seoul, South Korea}
}

@article{huo2014physiological,
  title={Physiological inverse tone mapping based on retina response},
  author={Huo, Yongqing and Yang, Fan and Dong, Le and Brost, Vincent},
  journal={The Visual Computer},
  volume={30},
  pages={507--517},
  year={2014},
  publisher={Springer}
}

@inproceedings{kovaleski2014high,
  title={High-quality reverse tone mapping for a wide range of exposures},
  author={Kovaleski, Rafael P and Oliveira, Manuel M},
  booktitle = {Proceedings of the SIBGRAPI Conference on Graphics, Patterns and Images},
  pages={49--56},
  year={2014},
  publisher = {IEEE},
  address = {Porto Alegre, Brazil}
}

@inproceedings{he2020conditional,
  title={Conditional sequential modulation for efficient global image retouching},
  author={He, Jingwen and Liu, Yihao and Qiao, Yu and Dong, Chao},
  booktitle={Proceedings of the European Conference on Computer Vision},
  pages={679--695},
  year={2020},
  address= {Glasgow, UK},
  publisher = {Springer}
}

@article{zeng2020learning,
  title={Learning image-adaptive 3d lookup tables for high performance photo enhancement in real-time},
  author={Zeng, Hui and Cai, Jianrui and Li, Lida and Cao, Zisheng and Zhang, Lei},
  journal={IEEE Transactions on Pattern Analysis and Machine Intelligence},
  volume={44},
  number={4},
  pages={2058--2073},
  year={2020},
  publisher={IEEE}
}

@inproceedings{zhang2012sr,
  title={SR-SIM: A fast and high performance IQA index based on spectral residual},
  author={Zhang, Lin and Li, Hongyu},
  booktitle={Proceedings of the IEEE International Conference on Image Processing},
  pages={1473--1476},
  year={2012},
  address = {San Diego, CA, USA},
  organization={IEEE}
}

@article{mantiuk2011hdr,
  title={HDR-VDP-2: A calibrated visual metric for visibility and quality predictions in all luminance conditions},
  author={Mantiuk, Rafa{\l} and Kim, Kil Joong and Rempel, Allan G and Heidrich, Wolfgang},
  journal={ACM Transactions on Graphics},
  volume={30},
  number={4},
  pages={1--14},
  year={2011},
  publisher={ACM New York, NY, USA}
}

@inproceedings{zhang2018unreasonable,
  title={The unreasonable effectiveness of deep features as a perceptual metric},
  author={Zhang, Richard and Isola, Phillip and Efros, Alexei A and Shechtman, Eli and Wang, Oliver},
  booktitle={Proceedings of the IEEE Conference on Computer Vision and Pattern Recognition},
  pages={586--595},
  year={2018},
  publisher = {IEEE},
  address = {Salt Lake City, UT, USA}
}

@article{liang2024vrt,
  title={Vrt: A video restoration transformer},
  author={Liang, Jingyun and Cao, Jiezhang and Fan, Yuchen and Zhang, Kai and Ranjan, Rakesh and Li, Yawei and Timofte, Radu and Van Gool, Luc},
  journal={IEEE Transactions on Image Processing},
  year={2024},
  publisher={IEEE}
}

@inproceedings{guo2023learning,
  title={Learning a practical {SDR-to-HDRTV} up-conversion using new dataset and degradation models},
  author={Guo, Cheng and Fan, Leidong and Xue, Ziyu and Jiang, Xiuhua},
  booktitle={Proceedings of the IEEE/CVF Conference on Computer Vision and Pattern Recognition},
  pages={22231--22241},
  address= {Vancouver, BC, Canada},
  year={2023}
}

@inproceedings{chen2021hdr,
	title = {{HDR} video reconstruction: A coarse-to-fine network and a real-world benchmark dataset},
	author={Chen, Guanying and Chen, Chaofeng and Guo, Shi and Liang, Zhetong and Wong, Kwan-Yee K and Zhang, Lei},
	booktitle={Proceedings of the IEEE/CVF International Conference on Computer Vision},
	pages={2502--2511},
	year={2021},
	address= {Montreal, QC, Canada}
}

@inproceedings{chung2023lan,
	title = {Lan-hdr: Luminance-based alignment network for high dynamic range video reconstruction},
	author={Chung, Haesoo and Cho, Nam Ik},
	booktitle={Proceedings of the IEEE/CVF International Conference on Computer Vision},
	pages={12760--12769},
	year={2023},
	address= {Paris, France},
}

@inproceedings{xu2024hdrflow,
	title = {Hdrflow: Real-time hdr video reconstruction with large motions},
	author={Xu, Gangwei and Wang, Yujin and Gu, Jinwei and Xue, Tianfan and Yang, Xin},
	booktitle={Proceedings of the IEEE/CVF Conference on Computer Vision and Pattern Recognition},
	pages={24851--24860},
	year={2024},
  address = {Seattle, WA, USA}
}

@inproceedings{chan2022basicvsr++,
	title = {Basicvsr++: Improving video super-resolution with enhanced propagation and alignment},
	author={Chan, Kelvin CK and Zhou, Shangchen and Xu, Xiangyu and Loy, Chen Change},
	booktitle={Proceedings of the IEEE/CVF Conference on Computer Vision and Pattern Recognition},
	pages={5972--5981},
	year={2022},
	address= {New Orleans, LA, USA},
}

@inproceedings{zhou2024upscale,
  title = {Upscale-a-video: Temporal-consistent diffusion model for real-world video super-resolution},
  author={Zhou, Shangchen and Yang, Peiqing and Wang, Jianyi and Luo, Yihang and Loy, Chen Change},
  booktitle={Proceedings of the IEEE/CVF Conference on Computer Vision and Pattern Recognition},
  pages={2535--2545},
  year={2024},
  address = {Seattle, WA, USA}
}

@inproceedings{ye2024deep,
  title={Deep Video Inverse Tone Mapping Based on Temporal Clues},
  author={Ye, Yuyao and Zhang, Ning and Zhao, Yang and Cao, Hongbin and Wang, Ronggang},
  booktitle={Proceedings of the IEEE/CVF Conference on Computer Vision and Pattern Recognition},
  pages={25995--26004},
  year={2024},
  address = {Seattle, WA, USA}
}

@article{zhou2021ibot,
  title={ibot: Image bert pre-training with online tokenizer},
  author={Zhou, Jinghao and Wei, Chen and Wang, Huiyu and Shen, Wei and Xie, Cihang and Yuille, Alan and Kong, Tao},
  journal={arXiv preprint arXiv:2111.07832},
  year={2021}
}

@inproceedings{wei2022masked,
  title={Masked feature prediction for self-supervised visual pre-training},
  author={Wei, Chen and Fan, Haoqi and Xie, Saining and Wu, Chao-Yuan and Yuille, Alan and Feichtenhofer, Christoph},
  booktitle={Proceedings of the IEEE/CVF Conference on Computer Vision and Pattern Recognition},
  pages={14668--14678},
  year={2022},
  address = {New Orleans, LA, USA}
}

@inproceedings{wang2023masked,
  title={Masked image modeling with local multi-scale reconstruction},
  author={Wang, Haoqing and Tang, Yehui and Wang, Yunhe and Guo, Jianyuan and Deng, Zhi-Hong and Han, Kai},
  booktitle={Proceedings of the IEEE/CVF Conference on Computer Vision and Pattern Recognition},
  pages={2122--2131},
  year={2023},
  address = {Vancouver, BC, Canada}
}

@inproceedings{liu2022learning,
  title={Learning quality-aware dynamic memory for video object segmentation},
  author={Liu, Yong and Yu, Ran and Yin, Fei and Zhao, Xinyuan and Zhao, Wei and Xia, Weihao and Yang, Yujiu},
  booktitle={Proceedings of the European Conference on Computer Vision},
  pages={468--486},
  year={2022},
  address = {Tel Aviv, Israel}
}

@inproceedings{pavlitska2023sparsely,
  title={Sparsely-gated mixture-of-expert layers for cnn interpretability},
  author={Pavlitska, Svetlana and Hubschneider, Christian and Struppek, Lukas and Z{\"o}llner, J Marius},
  booktitle={Proceedings of the IEEE International Joint Conference on Neural Networks},
  pages={1--10},
  year={2023},
  address = {Reno, NV, USA}
}

@inproceedings{teed2020raft,
  title={Raft: Recurrent all-pairs field transforms for optical flow},
  author={Teed, Zachary and Deng, Jia},
  booktitle={Proceedings of the European Conference on Computer Vision},
  pages={402--419},
  year={2020},
  address = {Gerald, France}
}

@article{tan2021deep,
  title={Deep sr-hdr: Joint learning of super-resolution and high dynamic range imaging for dynamic scenes},
  author={Tan, Xiao and Chen, Huaian and Xu, Kai and Jin, Yi and Zhu, Changan},
  journal={IEEE Transactions on Multimedia},
  volume={25},
  pages={750--763},
  year={2021},
  publisher={IEEE}
}

@article{jiang2023video,
  title={Video compression artifacts removal with spatial-temporal attention-guided enhancement},
  author={Jiang, Nanfeng and Chen, Weiling and Lin, Jielian and Zhao, Tiesong and Lin, Chia-Wen},
  journal={IEEE Transactions on Multimedia},
  volume={26},
  pages={5657--5669},
  year={2023},
  publisher={IEEE}
}

@article{han2022global,
  title={Global memory and local continuity for video object detection},
  author={Han, Liang and Yin, Zhaozheng},
  journal={IEEE Transactions on Multimedia},
  volume={25},
  pages={3681--3693},
  year={2022},
  publisher={IEEE}
}

@article{xiong2024rethinking,
  title={Rethinking Video Sentence Grounding From a Tracking Perspective With Memory Network and Masked Attention},
  author={Xiong, Zeyu and Liu, Daizong and Fang, Xiang and Qu, Xiaoye and Dong, Jianfeng and Zhu, Jiahao and Tang, Keke and Zhou, Pan},
  journal={IEEE Transactions on Multimedia},
  year={2024},
  publisher={IEEE}
}

@article{hong2022dual,
  title={Dual cross-attention for video object segmentation via uncertainty refinement},
  author={Hong, Jiahao and Zhang, Wei and Feng, Zhiwei and Zhang, Wenqiang},
  journal={IEEE Transactions on Multimedia},
  volume={25},
  pages={7710--7725},
  year={2022},
  publisher={IEEE}
}

@inproceedings{shu2024towards,
  title={Towards real-world HDR video reconstruction: A large-scale benchmark dataset and a two-stage alignment network},
  author={Shu, Yong and Shen, Liquan and Hu, Xiangyu and Li, Mengyao and Zhou, Zihao},
  booktitle={Proceedings of the IEEE/CVF Conference on Computer Vision and Pattern Recognition},
  pages={2879--2888},
  year={2024}
}

@inproceedings{zhu2024temporal,
  title={Temporal Residual Guided Diffusion Framework for Event-Driven Video Reconstruction},
  author={Zhu, Lin and Zheng, Yunlong and Zhang, Yijun and Wang, Xiao and Wang, Lizhi and Huang, Hua},
  booktitle={European Conference on Computer Vision},
  pages={411--427},
  year={2024},
  organization={Springer}
}

@inproceedings{zhu2022event,
  title={Event-based video reconstruction via potential-assisted spiking neural network},
  author={Zhu, Lin and Wang, Xiao and Chang, Yi and Li, Jianing and Huang, Tiejun and Tian, Yonghong},
  booktitle={Proceedings of the IEEE/CVF conference on computer vision and pattern recognition},
  pages={3594--3604},
  year={2022}
}

@inproceedings{zhu2021neuspike,
  title={Neuspike-net: High speed video reconstruction via bio-inspired neuromorphic cameras},
  author={Zhu, Lin and Li, Jianing and Wang, Xiao and Huang, Tiejun and Tian, Yonghong},
  booktitle={Proceedings of the IEEE/CVF international conference on computer vision},
  pages={2400--2409},
  year={2021}
}

\end{document}